\RequirePackage{iftex}\RequirePDFTeX
\documentclass[conference,10pt,letterpaper]{IEEEtran}
\usepackage{cite}
\usepackage{microtype}
\usepackage[hidelinks]{hyperref}
\hypersetup{
  pdftitle={Mind the Delta: Early Pruning of RTL Variants for Agentic Hardware Optimization},
  pdfauthor={Siyang Cai, Cangyuan Li, Haoyu Gao, Kun Wang, Mochen Li, Yilin Zhong, Ke Xu, Yinhe Han, Ying Wang}
}
\usepackage{booktabs}
\usepackage{graphicx}
\usepackage{amsmath,amssymb}
\usepackage{xcolor}
\usepackage{colortbl}
\usepackage{multirow}
\newsavebox{\tabbox}
\definecolor{oursblue}{HTML}{1F4E9C}
\colorlet{oursBg}{oursblue!20}
\colorlet{groupGray}{black!15}
\definecolor{refGray}{HTML}{8C8C8C}
\definecolor{barGray}{HTML}{D4708A}
\definecolor{stopRed}{HTML}{C8524B}
\usepackage{pifont}
\usepackage[scaled=0.92]{helvet}
\usepackage{tikz}
\usetikzlibrary{arrows.meta,positioning,shapes,calc}

\newcommand{\tabnote}[1]{\par\vspace{2pt}{\scriptsize\raggedright #1\par}}
\newcommand{\sysname}{\textsc{DiffScope}}
\renewcommand{\IEEEQED}{}%

\DeclareRobustCommand{\badge}[1]{\tikz[baseline=(b.base)]\node[circle, fill=oursblue, text=white, inner sep=0.5pt,
  minimum size=7.5pt, font=\tiny\bfseries] (b) {#1};}

\title{Mind the Delta: Early Pruning of RTL Variants\\for Agentic Hardware Optimization}
\author{%
  \IEEEauthorblockN{Siyang Cai\textsuperscript{1,2}, Cangyuan Li\textsuperscript{2},
    Haoyu Gao\textsuperscript{1,2}, Kun Wang\textsuperscript{2},
    Mochen Li\textsuperscript{3}}%
  \IEEEauthorblockN{Yilin Zhong\textsuperscript{3}, Ke Xu\textsuperscript{3},
    Yinhe Han\textsuperscript{2}, and Ying Wang\textsuperscript{2}}%
  \IEEEauthorblockA{%
    \textsuperscript{1}School of Advanced Interdisciplinary Sciences, University of Chinese Academy of Sciences, Beijing, China\\
    \textsuperscript{2}CICS, Institute of Computing Technology, Chinese Academy of Sciences, Beijing, China\\
    \textsuperscript{3}Sanechips Technology Co., Ltd., Shenzhen, China\\
    \texttt{caisiyang23@mails.ucas.ac.cn}\qquad\texttt{wangying2009@ict.ac.cn}}%
}

\begin{document}
\bstctlcite{IEEEexample:BSTcontrol}
\maketitle

\begin{abstract}
Autonomous coding agents powered by large language models (LLMs) now write diverse microarchitectural variants of existing register-transfer level (RTL) designs at high throughput, shifting the primary bottleneck of agentic RTL optimization onto logic synthesis. Commercial toolchains take hours per variant, creating an acute compute asymmetry between proposing and evaluating variants. Existing fast evaluators falter in this regime. Subtle RTL edits trigger nonlinear synthesis shifts that learned predictors and coding agents cannot resolve, while synthesis proxies at reduced effort either alter the optimization trajectory or retain most of its runtime. Selecting a variant, however, does not require regressing absolute quality of results across designs, only identifying when a candidate cannot surpass the best completed feasible implementation. Based on the observation that synthesis trajectories share strong design-specific bias with their base design, we propose \sysname{}, a training-free anytime selection framework that races full synthesis runs in parallel. By pairing live area telemetry with the already synthesized base trajectory, \sysname{} cancels early mapping inflation, reducing residual variance during initial mapping by 73.6$\times$, and stops uncompetitive runs via a lower bound calibrated on other design lineages. Across 43 design families from small blocks to processors with millions of cells, \sysname{} uniquely recovers the optimum of exhaustive synthesis in every family, cuts total evaluation work by 44.9\% (1.81$\times$), and reaches 2.18$\times$ on L-tier designs.
\end{abstract}

\section{Introduction}
\label{sec:intro}
Autonomous coding agents~\cite{claudecode, codex} powered by LLMs are changing how hardware designs are optimized. Given an existing RTL repository and a power, performance, and area (PPA) goal, they write dozens of functionally correct microarchitectural variants within minutes. Evaluating these variants has not become faster. Industrial flows judge each variant by the quality of results (QoR) from commercial logic synthesis such as Synopsys Design Compiler~\cite{dc}, since downstream implementation and sign-off start from its netlist. Such tools are black boxes whose internal algorithms users cannot modify, and a single run takes hours on a processor-scale design. This \emph{proposal--evaluation asymmetry} limits how many variants an agent can afford to explore.

\begin{figure}[t]\centering
{\definecolor{figWrong}{HTML}{C8524B}\definecolor{figOurs}{HTML}{D9E3F2}
\setlength{\tabcolsep}{1.05pt}\renewcommand{\arraystretch}{1.2}\fontsize{6.5}{7.6}\selectfont
\begin{tabular}{@{}l*{6}{c}cccc@{}}
\toprule
 & \multicolumn{2}{c}{\begin{tabular}{@{}c@{}}Learned\\[-0.5ex]predictor\end{tabular}} & \multicolumn{2}{c}{\raisebox{-0.2ex}{\includegraphics[height=1.4ex]{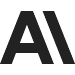}}\,Claude Code} & \multicolumn{2}{c}{\raisebox{-0.2ex}{\includegraphics[height=1.4ex]{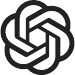}}\,Codex} & \begin{tabular}{@{}c@{}}Full\\[-0.5ex]synthesis\end{tabular} & \multicolumn{3}{c}{\textbf{\sysname{}}} \\
\cmidrule(lr){2-3}\cmidrule(lr){4-5}\cmidrule(lr){6-7}\cmidrule(l){9-11}
Design & GBDT & \begin{tabular}{@{}c@{}}Master\\[-0.5ex]RTL\end{tabular} & \begin{tabular}{@{}c@{}}Opus\\[-0.5ex]5.5\end{tabular} & \begin{tabular}{@{}c@{}}Sonnet\\[-0.5ex]5.5\end{tabular} & \begin{tabular}{@{}c@{}}GPT-6\\[-0.5ex]Astra\end{tabular} & \begin{tabular}{@{}c@{}}GPT-6.1\\[-0.5ex]Sol\end{tabular} & Time & Pick & Time & Speedup \\
\midrule
Gemmini 8$\times$8 & \textcolor{figWrong}{+4.2\%} & \textcolor{figWrong}{+1.8\%} & \textcolor{figWrong}{+1.8\%} & \textcolor{figWrong}{+4.2\%} & \textcolor{figWrong}{+5.5\%} & \textcolor{figWrong}{+5.5\%} & 6.96\,h & \cellcolor{figOurs}\ding{51} & \cellcolor{figOurs}2.99\,h & \cellcolor{figOurs}\textbf{2.3$\times$} \\
Vortex 1C & \textcolor{figWrong}{+62.8\%} & \ding{51} & \ding{51} & \ding{51} & \ding{51} & \textcolor{figWrong}{+62.8\%} & 5.70\,h & \cellcolor{figOurs}\ding{51} & \cellcolor{figOurs}1.67\,h & \cellcolor{figOurs}\textbf{3.4$\times$} \\
CVA6 & \textcolor{figWrong}{+25.6\%} & \ding{51} & \ding{51} & \textcolor{figWrong}{+25.6\%} & \textcolor{figWrong}{+29.0\%} & \textcolor{figWrong}{+29.0\%} & 2.77\,h & \cellcolor{figOurs}\ding{51} & \cellcolor{figOurs}1.01\,h & \cellcolor{figOurs}\textbf{2.7$\times$} \\
\midrule
Right picks & 0/3 & 2/3 & 2/3 & 1/3 & 1/3 & 0/3 & 3/3 & \multicolumn{3}{c}{\cellcolor{figOurs}\textbf{3/3}} \\
\bottomrule
\end{tabular}
}
\caption{A comparison of evaluators on picking the smallest-area timing-feasible variant among agent-written RTL candidates. Existing fast evaluators miss optimal picks. \sysname{} races full-synthesis runs, pruning uncompetitive candidates via base-anchored anytime bounds to achieve 100\% pick accuracy with 1.81$\times$ less evaluation work on 43 families.}
\label{fig:motiv}
\end{figure}

Agentic optimization therefore hinges on a selection problem. For each batch of variants an agent writes, the flow needs a fast way to judge their QoR and pick the best one. Existing evaluators make this choice without performing full synthesis. They fall into three groups. Coding agents estimate the QoR of each variant directly by reading its RTL. Learned predictors~\cite{masterrtl, rtltimer} train machine learning models on RTL features to predict QoR across designs. Synthesis proxies run a cheaper flow, such as open-source Yosys+ABC or a reduced-effort commercial mode, and rank variants by its results. All three groups fail on this task (Fig.~\ref{fig:motiv} and Table~\ref{tab:main}). Sibling variants differ by only a few percent in QoR, a gap finer than the error of learned or agent estimates and smaller than the distortion a cheaper flow introduces. None of them always finds the optimal design, and wrong picks raise the average area regret to as much as 33.8\% for coding agents.

Picking the best variant, however, does not require knowing the QoR of every variant. A variant can be dropped as soon as it clearly cannot beat the best feasible variant completed so far. From this perspective, a selector can launch full synthesis for all variants and stop the losing runs early, so the final pick still comes from full synthesis while the compute spent on losers shrinks. Designers already act on this intuition when they watch a long compile and kill a run that looks hopeless, and multi-fidelity pruners such as Successive Halving (SH)~\cite{karnin2013, jamieson2016} automate it by stopping runs on their intermediate ranks. Commercial synthesis logs make such ranks unreliable. Early logic expansion can inflate the intermediate area by up to 2.8$\times$ before restructuring and boundary optimization recover the final density, and the inflation differs from design to design. Intermediate ranks therefore invert, and rank-based pruning can discard the optimal variant before it recovers.

A closer look at the synthesis logs, however, shows that this inflation is not arbitrary. \emph{Variants derived from the same base design share its distortion profile}, because they keep most of its logic and hierarchy and pass through largely the same mapping steps. Since the base design is already synthesized, pairing a variant's live area log with the base log at the same compile phase cancels the shared inflation, so the remaining base-relative area delta reflects the variant's own edits. This pairing reduces the residual area variance during initial mapping by 73.6$\times$ (Table~\ref{tab:anchor}).

\begin{figure*}[t]\centering
\resizebox{.94\textwidth}{!}{\input{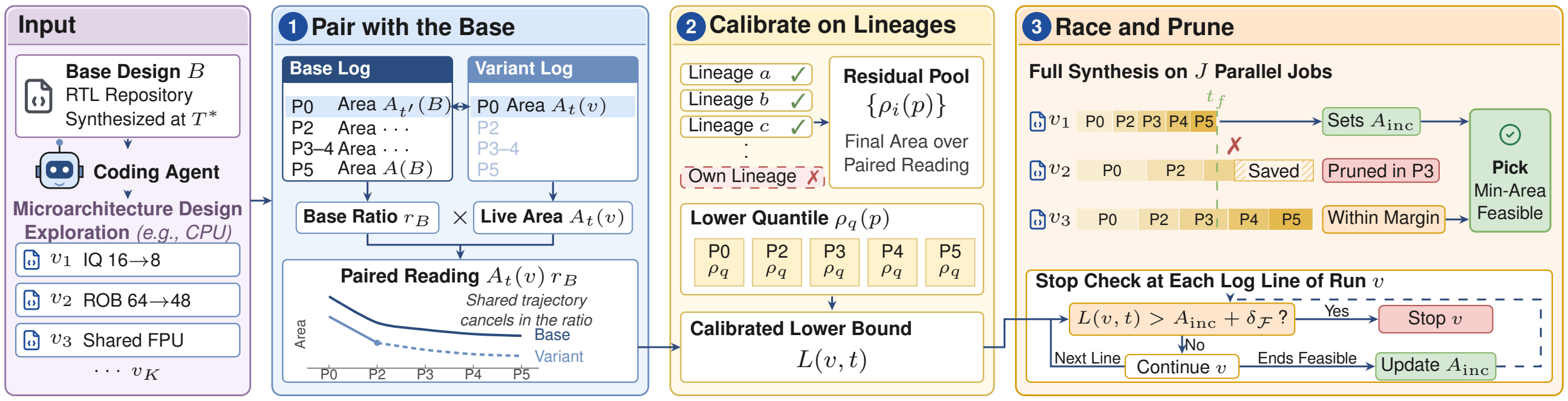}\unskip}
\caption{An overview of \sysname{} from agent-written variants to the selected design. The input is a base design $B$ already synthesized at $T^*$ and the variants $v_1,\dots,v_K$ that a coding agent writes from it. \badge{1} Each running variant's area $A_t(v)$ is paired with the base log at the same phase, and the base ratio $r_B$ cancels their shared trajectory. \badge{2} A per-phase lower quantile $\rho_q(p)$, calibrated with the family's own lineage held out, turns this reading into the calibrated lower bound $L(v,t)$. \badge{3} All variants run full synthesis on $J$ parallel jobs, a run stops once $L(v,t) > A_{\mathrm{inc}} + \delta_\mathcal{F}$, close runs complete, and the output is the minimum-area feasible variant.}
\label{fig:overview}
\end{figure*}

Building on this observation, we propose \sysname{}, a training-free anytime selection framework that races the full-synthesis runs of all variants in parallel (Fig.~\ref{fig:overview}). It first pairs each running variant's area log with the base log at the same compile phase, which cancels the shared inflation. It then turns this paired reading into a calibrated lower bound on the final area, using per-phase quantiles calibrated on other design lineages. Finally, it stops a run once this bound exceeds the best completed timing-feasible area plus a clock margin.

We construct a benchmark of 43 design families, each pairing a base design with six to eight variant versions of it, for 330 variants in total. Production coding agents wrote each variant toward a stated PPA goal and revised it over multiple turns until it passed functional tests. Across all 43 families, \sysname{} achieves \textbf{100\% strict top-1 accuracy} and always picks the same variant as exhaustive synthesis, which no evaluated baseline does. It cuts total evaluation work by 44.9\%, a 1.81$\times$ reduction that reaches 2.18$\times$ on Large-tier designs.

Our primary contributions are as follows.
\begin{itemize}
    \item We formalize agentic RTL variant selection as constrained best-arm identification and propose \sysname{}, which pairs each live synthesis log with its base design to cancel the shared inflation. \sysname{} stops uncompetitive runs inside an unmodified commercial synthesis flow, without training or tool changes.
    \item We calibrate one-sided lower bounds on other design lineages. They transfer to unseen designs, and every overshooting stop trails the incumbent by more than 40$\times$ its bound error.
    \item We build the first benchmark for QoR evaluation in agentic RTL optimization, with 43 design families written by production coding agents. We compare \sysname{} against 10 mainstream fast evaluators and schedulers on this benchmark, spanning learned predictors, coding agents, synthesis proxies, and multi-fidelity pruning, and show that its bounds transfer across clock targets without recalibration.
\end{itemize}

\section{Related Work}
\label{sec:relwork}
\textbf{Agentic RTL Optimization.} LLM-based and evolutionary optimizers~\cite{rtlrewriter, symrtlo, drrtl, veriopt, revolution, poet} rewrite RTL toward a PPA goal and use synthesis results as feedback, and AutoRTL-Bench~\cite{autortlbench} and RTL-OPT~\cite{rtlopt} benchmark such agents. These works have made proposing variants cheap and fast. However, they all leave evaluation unchanged, and each variant still costs one complete synthesis run. The bottleneck has therefore shifted from proposing variants to judging them.

\textbf{Fast QoR Evaluation.} Fast evaluators try to make judging cheaper. Learned predictors such as MasterRTL~\cite{masterrtl}, RTL-Timer~\cite{rtltimer}, RTL-Sequencer~\cite{rtlsequencer}, and CircuitFusion~\cite{circuitfusion} regress QoR from RTL features, and synthesis proxies run a cheaper flow such as Yosys~\cite{yosys} or a reduced-effort commercial mode. Related efforts learn QoR differences over high-level synthesis pragmas~\cite{diffhls} or accelerate resynthesis with custom engines~\cite{livesynth, smatch, liu2023aig}. These methods return an estimate in a fraction of the synthesis time. This speed comes at the cost of fidelity. Each estimate approximates commercial full-synthesis QoR, and its error often exceeds the few-percent gap between sibling variants of one design. Even the direction of an RTL change can flip across tools. RTL-OPT~\cite{rtlopt} finds that 24 of 43 human RTL optimizations improve QoR under Yosys but only 13 under Design Compiler. In industrial flows, sign-off starts from the commercial netlist, so only commercial QoR decides which variant is better.

\textbf{Early Termination and Multi-Fidelity Racing.} Early termination saves compute by stopping runs before they finish. In physical design, learned models forecast end-of-flow power from early placement metrics~\cite{doomedpower, kahngml}. Statistical racing~\cite{hoeffdingraces, evendar2006} stops candidates once confidence intervals from i.i.d.\ samples separate. Multi-fidelity pruners such as SH~\cite{karnin2013, jamieson2016}, Hyperband~\cite{hyperband}, and BOHB~\cite{bohb} keep the top-ranked runs at each budget. Even the non-stochastic analysis of SH~\cite{jamieson2016} relies on intermediate ranks converging to final ranks. In commercial synthesis, intermediate area inflates differently from design to design~\cite{chaos}, so ranks can reverse and stage-wise elimination can discard the winner. These methods judge a run only by its own early behavior rather than against a related design whose outcome is already known.

\section{Methodology}
\label{sec:method}
\sysname{} races the full-synthesis runs of candidate RTL variants in the three stages of Fig.~\ref{fig:overview}. It pairs each live area log with the base design $B$ (Sec.~\ref{sec:method-pair}), calibrates a lower bound on other design lineages (Sec.~\ref{sec:method-bound}), and stops uncompetitive runs against a feasible incumbent (Sec.~\ref{sec:method-incumbent}).

\subsection{Problem Formulation}
\label{sec:prob}
\textbf{Constrained Best-Arm Identification Model.} Let $V = \{v_1, \dots, v_K\}$ denote a set of $K$ functionally verified RTL variants written by an agent from base design $B$, which together form the family $\mathcal{F} = \{B\} \cup V$. Each variant $v$ is synthesized under a user-specified clock constraint $T^*$. Full synthesis reports final area $A(v)$ and worst negative slack $\mathrm{WNS}(v)$, with runtime $D_v$. A variant is timing-feasible if $\mathrm{WNS}(v) \ge 0$. Let $V^+ = \{v \in V \mid \mathrm{WNS}(v) \ge 0\}$ denote the feasible variant set. The selection objective is to identify the optimal variant:
\begin{equation}
v^* = \arg\min_{v \in V^+} A(v).
\end{equation}
If all variants violate timing ($V^+ = \emptyset$), the flow reports timing infeasibility and selects the variant minimizing slack violation. Unlike the i.i.d.\ samples of Hoeffding Races~\cite{hoeffdingraces} and the converging ranks of SH~\cite{jamieson2016}, compiler telemetry is deterministic, non-monotonic, and strongly autocorrelated, and timing closure is certified only upon completion.

\textbf{Evaluation Metrics and Operating Modes.} Selection quality is assessed by \emph{strict top-1 accuracy} ($\mathbb{I}[\hat{v} = v^*]$) and \emph{area regret} ($A(\hat{v})/A(v^*) - 1$). Computational efficiency is measured by total compile work $C = C_q + \sum_{v \in V} \tau_v$, where $\tau_v \le D_v$ is elapsed runtime before completion or pruning, and $C_q$ is candidate elaboration query cost, while the one-off base overhead $O$ is excluded. Parallel efficiency is evaluated by cluster makespan speedup over exhaustive synthesis. When an agent optimizes over multiple rounds, the base $B$ of each round is the previous round's winner, whose $T^*$ log is already on disk. Later rounds set $\delta_\mathcal{F} = 0$ to skip the $\pm 1\%$ clock runs, so $O = 0$ while candidate queries remain in $C_q$.

\subsection{Pairing with the Base Design}
\label{sec:method-pair}
\begin{table}[t]\centering
\caption{A comparison of the residual spread per compile phase with and without the base anchor over 43 families.}
\label{tab:anchor}
\resizebox{\columnwidth}{!}{%
\begin{tabular}{@{}lrrrrr@{}}
\toprule
Phase $p$ & $\mathrm{Var}(\epsilon_{\mathrm{raw}})$ & $\mathrm{Var}(\epsilon_{\mathrm{anchored}})$ & Var.\ ratio & Lower-tail $\rho_{0.01}$ & Lower-tail ratio \\
\midrule
P0 (initial mapping)   & 0.0537 & 0.00073 & 73.6$\times$ & 0.627 & 1.27$\times$ \\
P2 (boundary opt.)     & 0.0008 & 0.00223 & 0.36$\times$ & 0.887 & 1.05$\times$ \\
P3 (retiming/re-map)   & 0.0006 & 0.00101 & 0.59$\times$ & 0.942 & 1.03$\times$ \\
P4 (area recovery)     & 0.0002 & 0.00194 & 0.10$\times$ & 0.957 & 1.00$\times$ \\
P5 (late-stage opt.)   & 0.0001 & 0.00125 & 0.08$\times$ & 0.963 & 0.98$\times$ \\
\bottomrule
\end{tabular}}
\begin{minipage}{\columnwidth}\vspace{2pt}\tiny
Var.\ ratio divides raw residual variance by base-anchored variance (a ratio $>1$ indicates variance reduction). Lower-tail ratio compares 1st-percentile lower bounds with and without anchor. Each row covers the families where the phase occurs (P0 13, P4 38, and P2, P3, P5 all 43). Initial mapping (P0) accounts for 98.9\% of synthesis time saved, before queries.
\end{minipage}
\end{table}

Synthesis of $v$ passes through compile phases $p \in \mathcal{P} = \{\mathrm{P0}, \mathrm{P2}, \dots, \mathrm{P5}\}$, and its log reports intermediate area $A_t(v)$ at elapsed time $t$. Synthesis phases activate dynamically per design: P0 activates on large multi-level designs, namely all 12 L/XL families and one M family, P2, P3, and P5 activate on all 43 families, and P4 activates on 38 families. Phase P1 occurs in only two families and is omitted. We model $A_t(v) = A(v) \cdot g_\mathcal{F}(p, t) \cdot \exp(\epsilon_v(p, t))$, where trajectory factor $g_\mathcal{F}$ captures compiler mapping inflation of family $\mathcal{F}$ and $\epsilon_v$ is the residual error. Because variant $v$ is an agentic mutation of base $B$, their trajectories share the dominant term $g_\mathcal{F}$. Taking the ratio to the base trajectory approximately cancels $g_\mathcal{F}$:
\begin{equation}
\label{eq:ratio}
\frac{A_t(v)}{A_{t'}(B)} = \frac{A(v)}{A(B)} \cdot \exp(\epsilon_v - \epsilon_B),
\end{equation}
where $t'$ is the base reading at the same elapsed time since entering phase $p$, clamped to the base phase end. The base log is complete before the race, so pairing reads only the log prefix of $v$, and a phase missing from the base triggers no stop. A phase-median base ratio keeps all 43 picks and changes total work by $<0.1\%$. Solving for $A(v)$ yields:
\begin{equation}
\label{eq:resid}
A(v) = A_t(v) \cdot \frac{A(B)}{A_{t'}(B)} \cdot \exp(\epsilon_B - \epsilon_v).
\end{equation}
Here $r_B = A(B)/A_{t'}(B)$ is the base ratio, and $A_t(v)\,r_B$ is the paired reading of $v$ at phase $p$.

Table~\ref{tab:anchor} follows from this paired formulation. During initial mapping (P0), structural gate inflation ($g_{\mathcal{F}} \gg 1$) dominates the raw error, and base anchoring cancels this shared bias. In later phases (P2--P5), restructuring is nearly complete ($g_{\mathcal{F}} \approx 1$), so dividing by base telemetry adds $\mathrm{Var}(\epsilon_B)$ and the two-sided variance ratio falls below 1.0$\times$. Pruning safety, however, depends only on the one-sided lower tail. Anchoring tightens its 1st percentile by 1.27$\times$ in P0 and keeps it within 5\% of the raw tail in later phases. The three runs whose bound overshoots final area by more than $\delta_\mathcal{F}$ overshoot by at most 0.48\%, yet each trails the incumbent by at least 26\%, more than 40$\times$ its overshoot in absolute area. No optimal variant was stopped in our runs.

\subsection{Calibrating Lower Bounds on Lineages}
\label{sec:method-bound}
By Eq.~(\ref{eq:resid}), the final area of $v$ equals its paired reading times the residual factor $\rho = A(v)/(A_t(v)\,r_B)$. \sysname{} collects $\rho$ at phase $p$ from the calibration lineages into a residual pool $\{\rho_i(p)\}$, where the calibration set always excludes the lineage of $v$. Given quantile parameter $q$, it computes $\rho_q(p) = \mathrm{Quantile}_q(\{\rho_i(p)\}_{i \in \mathrm{calib}})$. The calibrated lower bound at elapsed time $t$ in phase $p$ is:
\begin{equation}
\label{eq:bound}
L(v,t) = A_t(v) \cdot r_B \cdot \rho_q(p).
\end{equation}

\subsection{Racing and Pruning with a Feasible Incumbent}
\label{sec:method-incumbent}
\sysname{} tracks the incumbent feasible variant $v_{\mathrm{inc}} = \arg\min_{v \in V_{\mathrm{done}}^+} A(v)$ (with area $A_{\mathrm{inc}} = A(v_{\mathrm{inc}})$), where $V_{\mathrm{done}}^+ = \{v \in V^+ \mid v \text{ has completed}\}$. \emph{Only runs that finish full synthesis with non-negative slack update $v_{\mathrm{inc}}$}. Compact variants that fail timing ($\mathrm{WNS} < 0$) are rejected and never update $A_{\mathrm{inc}}$, preventing timing-infeasible runs from falsely tightening the pruning threshold.

To establish an incumbent early when $J \ge 1$ parallel synthesis jobs are available, variants are dispatched in ascending order of estimated complexity. Prior to synthesis, a lightweight elaboration query reports the elaborated leaf-cell count $N_v$, which takes a median of 7.8\,s per variant and about 1.5\,h on the largest XL design, and its time is charged to the cost of \sysname{}. Smaller variants compile faster, so an incumbent arrives early. Ordering by file name needs no query and still keeps 100\% top-1 accuracy (Sec.~\ref{sec:eval-abl}).

At every log line, a run is stopped once its lower bound exceeds the feasible incumbent plus a clock margin:
\begin{equation}
\label{eq:stop}
L(v,t) > A_{\mathrm{inc}} + \delta_\mathcal{F},
\end{equation}
where $\delta_\mathcal{F} = \max_{c \in \{0.99, 1.01\}} |A(B;\, c\,T^*) - A(B;\, T^*)|$ is the clock margin, measured by synthesizing the base at $\pm 1\%$ clock targets.

\subsection{Conditional Soundness and Operating Regimes}
\label{sec:guarantee}
\textbf{Soundness of Stopping.} Let $v_{\mathrm{inc}} \in V^+$ be an established feasible incumbent, and let variant $v$ be stopped at time $t$ under Eq.~(\ref{eq:stop}). If the bound covers $v$, i.e., $A(v) \ge L(v,t)$, then $A(v) > A(v_{\mathrm{inc}})$, so $v$ cannot be the unique optimal variant $v^*$. This follows directly from the stopping rule. A stop implies $L(v,t) > A_{\mathrm{inc}} + \delta_\mathcal{F} \ge A(v_{\mathrm{inc}})$, and coverage gives $A(v) \ge L(v,t) > A(v_{\mathrm{inc}})$. Since $v_{\mathrm{inc}} \in V^+$, variant $v$ is strictly worse in area than a known feasible implementation.

\textbf{Regret Transfer.} Suppose that, for the optimal variant $v^*$, the bound holds at every check of phase $p$ with probability at least $1-q$. Then \sysname{} stops $v^*$ with probability at most $|\mathcal{P}|\,q$, or 2.5\% at the default $q = 0.005$. Checks along one log are correlated, so per-line quantiles do not ensure this condition. We therefore treat it as an assumption rather than a guarantee and check it empirically on all stop events and picks (Sec.~\ref{sec:risk}).

\textbf{Operating Regimes and Fallback Dynamics.} Until a feasible run completes, $A_{\mathrm{inc}} = +\infty$ keeps Eq.~(\ref{eq:stop}) inactive, so if $V^+ = \emptyset$ the flow reduces to exhaustive synthesis. For standalone deployment without prior logs, setting $\delta_\mathcal{F} = 0$ avoids auxiliary $\pm 1\%$ clock runs while preserving zero selection regret (Sec.~\ref{sec:eval-pool}).

\section{Experiments}
\label{sec:experiments}

\begin{table}[t]
\caption{Statistics of the benchmark per size tier over 43 families.}
\label{tab:bench}
\centering
{\footnotesize
\setlength{\tabcolsep}{3pt}
\resizebox{\columnwidth}{!}{%
\begin{tabular}{l|rrrlrrr}
\toprule
Tier & Fam. & Var. & $N_B$ range & Example axes & $T^*$ time & Gap & Viol. \\
\midrule
S & 23 & 184 & 0.9K--15K & TDM/parallel, radix & 71\,s & 12.4\% & 2 \\
M & 8 & 64 & 25K--170K & MulDiv unroll, queues & 233\,s & 5.0\% & 5 \\
L & 10 & 70 & 346K--959K & Issue queue, ROB, FPU & 51\,min & 2.3\% & 0 \\
XL & 2 & 12 & 2.27M--3.29M & FMA sharing, buffers & 6.4\,h & 2.1\% & 0 \\
\midrule
All & 43 & 330 & 0.9K--3.29M & -- & 95\,s & 7.1\% & 7 \\
\bottomrule
\end{tabular}%
}
}

\tabnote{Design tiers are classified by the mapped cell count $N_B$ of the base at $T^*$. Target clock runtimes reflect median variant compile time at $T^*$. Area gap indicates median percentage area deviation between candidate variants and base, and timing violations count candidates failing $T^*$.}
\end{table}

\begin{table*}[t]
\caption{A comparison of all evaluators on pick quality and synthesis cost over 43 families. Top-1 and regret are in percent.}
\label{tab:main}
\centering
\sbox{\tabbox}{{\footnotesize
\setlength{\tabcolsep}{3.4pt}
\renewcommand{\arraystretch}{0.92}
\begin{tabular}{lllcccccrl}
\toprule
 & & & \multicolumn{4}{c}{Strict top-1 ($\uparrow$)} & Feas.-pick & \multicolumn{2}{c}{} \\
\cmidrule(lr){4-7}
Type & Decision & Evaluator & All (43) & {\scriptsize S (23)} & {\scriptsize M (8)} & {\scriptsize L/XL (12)} & regret ($\downarrow$) & \multicolumn{2}{c}{Speedup ($\uparrow$)} \\
\midrule
\multirow{4}{*}{Learned predictor} & \multirow{2}{*}{Own pick} & MasterRTL (rebuilt, zero-shot)~\cite{masterrtl} & 44 & {\scriptsize 43} & {\scriptsize 75} & {\scriptsize 25} & 25.17 & 4.09$^*$ & \textcolor{barGray}{\rule{6.2mm}{1.1ex}}\hspace{0.25mm}\tikz[baseline=0pt]\fill[barGray] (0,0)--(0,1.1ex)--(0.75ex,0.55ex)--cycle; \\
 &  & GBDT on operator counts & 21 & {\scriptsize 30} & {\scriptsize 25} & {\scriptsize 0} & 29.21 & 14.56$^*$ & \textcolor{barGray}{\rule{6.2mm}{1.1ex}}\hspace{0.25mm}\tikz[baseline=0pt]\fill[barGray] (0,0)--(0,1.1ex)--(0.75ex,0.55ex)--cycle; \\
\cmidrule(l){2-10}
 & \multirow{2}{*}{Top-$k$ + full synth.} & GBDT on operator counts & 88 & {\scriptsize \textbf{100}} & {\scriptsize \textbf{100}} & {\scriptsize 58} & \textbf{0.00} & 0.94$^*$ & \textcolor{barGray}{\rule{3.64mm}{1.1ex}} \\
 &  & MasterRTL (rebuilt, zero-shot) & 84 & {\scriptsize 96} & {\scriptsize \textbf{100}} & {\scriptsize 50} & \textbf{0.00} & 0.80$^*$ & \textcolor{barGray}{\rule{3.09mm}{1.1ex}} \\
\midrule
\multirow{6}{*}{Synthesis proxy} & \multirow{3}{*}{Own pick} & Yosys+ABC+OpenSTA & 58 & {\scriptsize 74} & {\scriptsize 75} & {\scriptsize 17} & 2.85 & 1.06$^*$ & \textcolor{barGray}{\rule{4.10mm}{1.1ex}} \\
 &  & Commercial, medium effort & 77 & {\scriptsize 87} & {\scriptsize 88} & {\scriptsize 50} & 1.14 & 0.88 & \textcolor{barGray}{\rule{3.40mm}{1.1ex}} \\
 &  & Commercial, hierarchy kept & 77 & {\scriptsize \textbf{100}} & {\scriptsize 88} & {\scriptsize 25} & 0.01 & 1.02 & \textcolor{barGray}{\rule{3.94mm}{1.1ex}} \\
\cmidrule(l){2-10}
 & \multirow{3}{*}{Top-$k$ + full synth.} & Yosys+ABC+OpenSTA & 77 & {\scriptsize 91} & {\scriptsize \textbf{100}} & {\scriptsize 33} & 0.04 & 0.48$^*$ & \textcolor{barGray}{\rule{1.86mm}{1.1ex}} \\
 &  & Commercial, medium effort & 95 & {\scriptsize 96} & {\scriptsize \textbf{100}} & {\scriptsize 92} & 0.12 & 0.49 & \textcolor{barGray}{\rule{1.90mm}{1.1ex}} \\
 &  & Commercial, hierarchy kept & 79 & {\scriptsize \textbf{100}} & {\scriptsize 88} & {\scriptsize 33} & $<$0.01 & 0.57 & \textcolor{barGray}{\rule{2.20mm}{1.1ex}} \\
\midrule
\multirow{4}{*}{Coding agent} & \multirow{4}{*}{Own pick} & Claude Opus 5.5 (Claude Code) & 63 & {\scriptsize 65} & {\scriptsize 75} & {\scriptsize 50} & 8.31 & -- &  \\
 &  & GPT-6 Astra (Codex) & 58 & {\scriptsize 65} & {\scriptsize 62} & {\scriptsize 42} & 6.43 & -- &  \\
 &  & GPT-6.1 Sol (Codex) & 47 & {\scriptsize 52} & {\scriptsize 50} & {\scriptsize 33} & 13.55 & -- &  \\
 &  & Claude Sonnet 5.5 (Claude Code) & 37 & {\scriptsize 39} & {\scriptsize 12} & {\scriptsize 50} & 33.81 & -- &  \\
\midrule
Multi-fidelity & Stage pruning & Successive halving (SH)$^\dagger$~\cite{karnin2013,jamieson2016} & 98 & {\scriptsize \textbf{100}} & {\scriptsize \textbf{100}} & {\scriptsize 92} & 0.01 & 1.12 & \textcolor{barGray}{\rule{4.33mm}{1.1ex}} \\
\midrule
\rowcolor{oursBg}\textbf{Ours} & \textbf{Early pruning} & \textbf{\sysname{}} & \textbf{100} & \textbf{100} & \textbf{100} & \textbf{100} & \textbf{0.00} & \textbf{1.81} & \textcolor{oursblue}{\rule{7.00mm}{1.1ex}} \\
\midrule
\multicolumn{3}{l}{Reference: full synthesis of all $K$ variants} & 100 & 100 & 100 & 100 & 0 & 1.00 & \textcolor{barGray}{\rule{3.87mm}{1.1ex}} \\
\bottomrule
\end{tabular}
}
}\usebox{\tabbox}
\par\vspace{2pt}\parbox{\wd\tabbox}{\scriptsize\raggedright An own pick is the variant an evaluator ranks best, returned without full synthesis. A top-$k$ decision fully synthesizes the $k$ best-ranked variants and returns the smallest feasible one. Top-1 counts families whose pick is the exhaustive optimum, and a family without an output counts as a miss. Regret is the mean area excess $A(\hat{v})/A(v^*)-1$ over feasible picks. Speedup divides exhaustive synthesis work by the evaluator's total work. Front-end failures and the 8\,h cap (Yosys on 4 designs) leave some families without complete cost, so $^*$ speedups cover only cost-complete families (10 of 36 for MasterRTL). $^\dagger$ marks arrival-time replay, and \tikz[baseline=0pt]\fill[barGray] (0,0)--(0,1.1ex)--(0.75ex,0.55ex)--cycle; marks bars beyond the $1.81\times$ cap. Coding-agent runtime varies with network and API latency and is not reported.}
\end{table*}

\begin{figure*}[t]
\centering
\begin{minipage}[t]{4.08in}\centering
\begin{minipage}[t]{2.50in}\centering \includegraphics{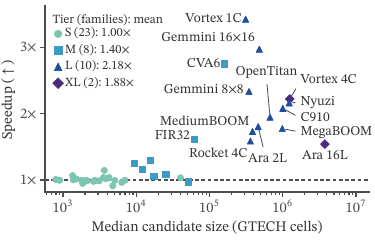}\par\centerline{\scriptsize (a) Per family vs.\ candidate size ($J{=}1$)}\end{minipage}\hfill
\begin{minipage}[t]{1.50in}\centering \includegraphics{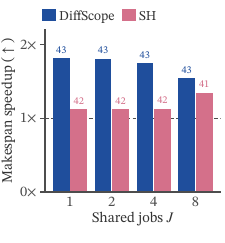}\par\centerline{\scriptsize (b) Shared job pool}\end{minipage}
\caption{A breakdown of synthesis time saved by \sysname{} on 43 families. The saving grows with design size up to L-tier and holds when all selections share one job pool. (a) Speedup vs.\ median candidate size, with per-family tier means in the legend. (b) Makespan speedup vs.\ $J$, labeled with correct picks of 43$^\dagger$.}
\label{fig:scaling}
\end{minipage}\hfill
\begin{minipage}[t]{2.93in}\centering
\begin{minipage}[t]{1.42in}\centering \includegraphics{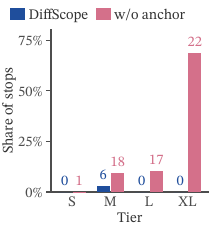}\par\centerline{\scriptsize (a) Overshooting stops per tier}\end{minipage}\hfill
\begin{minipage}[t]{1.47in}\centering \includegraphics{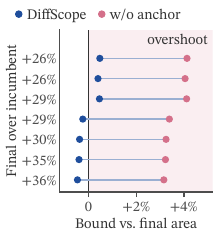}\par\centerline{\scriptsize (b) CVA6 stops at $J{=}1$}\end{minipage}
\caption{A stop-safety audit of \sysname{} with and without the base anchor. Base anchoring confines worst-case CVA6 overshoot to +0.48\% while stopped runs trail by +26\% to +36\%. (a) Share of stops whose bound exceeds final area by $>\delta_\mathcal{F}$ over $J\in\{1,2,4,8\}$, with counts on bars. (b) Bound error of stopped CVA6 runs at $J{=}1$ vs.\ final area over incumbent.}
\label{fig:risk}
\end{minipage}
\end{figure*}

\subsection{Experimental Setup}
\label{sec:eval-bench}\label{sec:eval-setup}
\textbf{Benchmark.} We construct a diverse, functionally verified benchmark of 43 design families built on 39 open-source projects and the RTLLM and RTL-OPT benchmarks~\cite{rtllm, rtlopt} (Table~\ref{tab:bench}). Its 330 agent-written variants fall into four tiers, Small (S), Medium (M), Large (L), and Extra-Large (XL), by base cell count $N_B$, from 0.9K-cell blocks to the 3.29M-cell 16-lane Ara vector processor~\cite{ara2} and Vortex-4~\cite{vortex}. Other families build on BOOM~\cite{chipyard}, Rocket~\cite{rocket}, Gemmini~\cite{gemmini}, CVA6~\cite{cva6}, OpenC910~\cite{c910}, Nyuzi (formerly Nyami)~\cite{nyuzi}, OpenTitan Earl Grey~\cite{opentitan}, VeeR EH2~\cite{veer}, NVDLA~\cite{nvdla}, VexRiscv~\cite{d-vexriscv}, and PicoRV32~\cite{d-picorv32}.

\textbf{Construction rules.} Each family starts from an existing base repository and one optimization goal, either reducing area or closing a tighter clock. Given the base and the goal, Codex~\cite{codex} with GPT-6.1 Sol and Claude Code~\cite{claudecode} with Claude Opus 5.5 write six to eight variants per family over multiple turns. Every family fixes two microarchitectural axes, such as the issue queue and reorder buffer of a processor or the multiplier unrolling of an accelerator (Table~\ref{tab:bench}). Its variants cover alternatives along each axis, their combinations, and free-form rewrites toward the goal. Each variant changes state bits, iteration counts, or word-level structure, so the variants are distinct microarchitectural choices rather than coding-style edits. The families span Verilog (28), SystemVerilog (7), Chisel (6), and SpinalHDL (2), and the last two enter every tool as generated Verilog. A family is admitted only if its testbench catches at least three injected faults, every variant passes it and synthesizes cleanly, and a manual check of full-synthesis results rejects only families whose variants yield identical QoR. Pre-screening uses elaboration statistics only, so no evaluated proxy biases the benchmark. Full synthesis runs Synopsys Design Compiler~\cite{dc} in its high-effort \texttt{compile\_ultra} mode, targeting the NanGate 45\,nm Open Cell Library under conservative operating conditions. We use Design Compiler as a testbed rather than a dependency. \sysname{} reads only the intermediate area that the tool logs at each compile phase, and the base ratio $r_B$ is dimensionless, so any synthesis tool that reports phased QoR can host it. The 43 families form 28 lineage groups, and all quantiles are calibrated leave-one-lineage-out to test generalization to unseen RTL lineages.

\textbf{Baselines.} We compare \sysname{} against 10 fast evaluators in four groups.
\begin{enumerate}
    \item \textbf{Learned predictors.} MasterRTL~\cite{masterrtl} is rebuilt and trained on 60 external designs disjoint from our families, where it reaches a cross-design Spearman correlation of 0.95. It is evaluated zero-shot on our families, together with gradient boosted decision trees (GBDT) fit on elaboration operator counts. Their top-$k$ setting further grants full synthesis of the best-ranked candidates.
    \item \textbf{Synthesis proxies.} An open-source flow uses Yosys+ABC+OpenSTA, and two fast Design Compiler modes run at medium optimization effort and with preserved hierarchical boundaries.
    \item \textbf{Coding agents.} Claude Opus 5.5 and Claude Sonnet 5.5 in Claude Code and GPT-6 Astra and GPT-6.1 Sol in Codex judge variants zero-shot from candidate RTL diffs, the target clock, and base synthesis logs. Each agent evaluates a family in a fresh session with a clean context and an isolated workspace. It can read files and run shell commands but cannot launch full commercial synthesis, since that would reduce it to exhaustive synthesis.
    \item \textbf{Multi-fidelity schedulers.} SH~\cite{karnin2013, jamieson2016} is adapted to synthesis phases and keeps the top 50\% of runs at P2, P3, and P4.
\end{enumerate}

\textbf{Cost accounting.} Evaluator cost includes all overheads, namely generic-technology (GTECH) elaboration queries ($C_q$), proxy runtimes, and the synthesis prefixes consumed by SH. Queries total 15.7\,h, 9.1\,h of it on the 3.29M-cell Ara 16L. With parallel jobs, a slot stays occupied for $\min(D_v, t_{\mathrm{stop}}(v) + \Delta_v)$, where $\Delta_v$ is the termination latency. Other results replay recorded logs, exact up to timestamps since Design Compiler is deterministic. Online runs on axi\_dma (S), CVA6 (M), and Vortex 1C (L) at $J \in \{1, 4\}$ test stops under load, including P0 stops on Vortex 1C, stopping runs via \texttt{SIGTERM}. They reproduce the replayed picks in all six runs and stop sets in all six, with work within 2.3\% of replay and a median termination latency of 0.8\,s.

\subsection{Pick Quality and Cost}
\label{sec:eval-comp}
\sysname{} is the only evaluator that picks the exhaustive winner in all 43 families, and it does so at a 1.81$\times$ speedup (Table~\ref{tab:main}). Competing evaluators fail in three distinct ways.

First, learned predictors capture cross-design scale but collapse on variants of the same base design. GBDT and MasterRTL reach global Spearman correlations of 0.86 and 0.95, yet their median within-family Kendall's $\tau_b$ drops to 0.26 and 0.62. Their median error exceeds the top-2 variant gap in 30/38 and 25/36 families, so cross-validation selects $k=8$ and recovers accuracy only by synthesizing every candidate. Frontier coding agents face the same wall. The best one, Claude Opus 5.5, reaches only 63\% top-1 with 8.31\% regret, and the four agents together mispredict 30 of their 46 scored picks in the 16 families with a $<2\%$ gap.

Second, synthesis proxies rank variants well but cannot save work safely. Their median within-family $\tau_b$ is 0.93--1.00, yet the medium-effort and hierarchy-preserving modes still miss 23\% of picks. They give only 1.02$\times$ with preserved hierarchy and slow down to 0.49--0.57$\times$ under top-$k$. Yosys+ABC drops to 58\% top-1 with an 11.2\% median regret on its misses and hits the 8\,h cap on four large processors.

Third, stage-based SH reaches 98\% top-1 at only 1.12$\times^\dagger$. It misses OpenC910~\cite{c910} because it discards the lower half by rank at fixed phases, before the final boundary optimization lets the winner recover. \sysname{} never ranks runs and stops one only when its calibrated bound exceeds a completed feasible area, so close runs complete.

\begin{table}[t]
\caption{An ablation study of \sysname{} components over all 43 families at $J{=}1$. Top-1 is in percent.}
\label{tab:abl}
\centering
{\footnotesize
\setlength{\tabcolsep}{3.2pt}
\renewcommand{\arraystretch}{0.95}
\begin{tabular}{@{}lcccc@{}}
\toprule
 & Top-1 & Synthesis & Query & \\
Configuration & ($\uparrow$) & (h, $\downarrow$) & (h, $\downarrow$) & Speedup ($\uparrow$) \\
\midrule
\rowcolor{oursBg}\textbf{\sysname{}} & \textbf{100} & 70.4 & 15.7 & \textbf{1.81$\times$} \\
\quad w/o base anchor & \textcolor{stopRed}{\textbf{98}} & 95.1 & 15.7 & 1.41$\times$ \\
\quad w/o size order (name order) & 100 & 91.9 & 0 & 1.70$\times$ \\
\quad fixed margin, no calibration & 100 & 87.1 & 15.7 & 1.52$\times$ \\
\quad w/o anchor, oracle $q$ & 100 & 116.8 & 15.7 & 1.18$\times$ \\
\quad w/o anchor, nested $q$ & \textcolor{stopRed}{\textbf{98}} & 108.2 & 15.7 & 1.26$\times$ \\
\quad anchor in P0 only & 100 & 72.1 & 15.7 & 1.78$\times$ \\
\midrule
\textit{Reference: final-area order} & \textit{100} & \textit{69.1} & \textit{15.7} & \textit{1.84$\times$} \\
\bottomrule
\end{tabular}
}

\tabnote{Red marks configurations that mispredict. Speedup represents exhaustive synthesis time (156.3\,h) divided by combined synthesis and query time. Final-area ordering provides an unavailable post-synthesis oracle reference with equal query time.}
\end{table}

\subsection{Where the Savings Come From}
\label{sec:eval-scale}\label{sec:eval-abl}
Savings grow with design size (Fig.~\ref{fig:scaling}(a)), because uncompetitive variants trail early in long logic restructuring. S-tier blocks (median 71\,s compile) are dominated by elaboration and tool startup and see $1.00\times$. Fifteen families, 14 of them S-tier, fall slightly below $1\times$, at worst $0.91\times$ on crc32. Speedup climbs to $1.40\times$ on M (median 233\,s), $2.18\times$ on L (median 51\,min), and $1.88\times$ on XL (median 6.4\,h). L and XL families supply 96.6\% of all savings, L families save a mean net 3.59\,h each after queries, and Vortex 1C reaches $3.4\times$. XL falls slightly below L for two reasons. XL variants lie within a median 2.1\% of base area (Table~\ref{tab:bench}), so more stay inside the $\delta_{\mathcal{F}}$ band and complete to preserve zero regret. The incumbent time $t_f = \min_{v \in V^+} (s_v + D_v)$, with dispatch time $s_v$, also grows on multi-million-cell cores, narrowing the stopping window.

Table~\ref{tab:abl} isolates each component. Without the base anchor, \sysname{} drops to 98\% top-1 by mispredicting Ara 2L with an absolute area loss of 7,917\,$\mu\mathrm{m}^2$, and its speedup falls to $1.41\times$. An oracle $q$ tuned on the test families avoids this miss but leaves only $1.18\times$, and a $q$ chosen by nested leave-one-lineage-out selection still mispredicts one family, keeping 42 of 43 picks at 1.26$\times$. Anchoring at P0 alone gives $1.78\times$, so cancelling P0 mapping inflation supplies most of the savings. A fixed uncalibrated margin in place of lineage quantiles keeps all picks but drops to $1.52\times$, since it must be conservative for every design. Without size ordering, speedup falls to $1.70\times$ with 91.9\,h of synthesis and no query cost. GTECH ordering lets a small feasible variant finish first and cuts synthesis by 21.5\,h, within 1.9\% of the final-area oracle at 69.1\,h, which outweighs the 15.7\,h query cost and lifts speedup to $1.81\times$.

\subsection{Stop Safety}
\label{sec:risk}
We audit all $942$ stop events across $172$ trajectories at $J \in \{1,2,4,8\}$ (Fig.~\ref{fig:risk}). Only six stops, from three M-tier CVA6 runs replayed across $J$, overshoot, i.e., exceed final area by more than $\delta_{\mathcal{F}}$ (Fig.~\ref{fig:risk}(a)). Their worst overshoot is $+0.48\%$ of final area, whereas these variants trail the incumbent by $+26\%$ to $+29\%$, more than 40$\times$ their overshoot in absolute area (Fig.~\ref{fig:risk}(b)). In all 16 families whose top two variants differ by $<2\%$, \sysname{} picks the winner and lets 15 runners-up complete. Without the base anchor, $58/903$ stops overshoot (up to $18.8\%$) and three harmful stops hit Ara 2L. Across 181 sensitivity settings over $q \in [0.001, 0.05]$, lineage splits, and $J$, \sysname{} keeps $43/43$ picks at $1.81\text{--}1.95\times$.

\textbf{Tighter clocks.}\label{sec:tight-clock} If the clock tightens between rounds, re-synthesizing the base at $T^*_{\mathrm{new}}$ captures the shift, so $\rho_q(p)$ needs no recalibration. On 13 M--XL families at the tightest base-feasible $T^*_{\mathrm{new}} \in \{0.97, 0.94, 0.90, 0.85\}T^*$, where $6/102$ variants fail timing, frozen quantiles keep $13/13$ picks at every $J$ and reach $1.93\times$ at $J{=}1$. Unanchored bounds at $J{=}1$ and SH at $J \ge 2^\dagger$ both mispredict. If any completed run could set $A_{\mathrm{inc}}$ regardless of WNS (Sec.~\ref{sec:method-incumbent}), a smaller infeasible variant would stop the feasible winner in two families, leaving 11/13 picks.

\subsection{Deployment Robustness}
\label{sec:eval-pool}
A single family run with $J \ge K$ starts all variants before the first one finishes at $t_f = \min_{v \in V^+} (s_v + D_v)$, so few runs remain to stop and net speedup falls to $0.95\times^\dagger$ after queries. A breadth-first pool shared across families staggers completions and removes this cold-incumbent window (Fig.~\ref{fig:scaling}(b)). On the shared pool, \sysname{} keeps $43/43$ picks at $1.75\text{--}1.81\times$ for $J \le 4$ and $1.55\times$ at $J=8$. SH reaches only $1.12\text{--}1.35\times$ and mispredicts one family at $J \le 4$ and two at $J=8$. The gain thus lies in total work and shared-pool throughput, not single-family latency. With $\delta_\mathcal{F} = 0$, skipping $\pm 1\%$ clock runs, \sysname{} keeps $43/43$ picks at the same $1.81\times$ from one $T^*$ base log. Charging initial base synthesis ($65.3$\,h) to both sides and a one-off $1.1$\,h of anchor preparation to \sysname{} in round 1, and assuming equal cost per round, cumulative speedup grows from 1.45$\times$ in round 1 to 1.58$\times$, 1.68$\times$, and 1.74$\times$ by rounds 2, 4, and 8, approaching the $1.81\times$ limit.

\section{Conclusion}
\label{sec:conclusion}
We propose \sysname{}, a training-free anytime selection framework for agentic RTL optimization. It pairs each live synthesis log with the base design and stops a run once a lineage-calibrated bound shows that it cannot win. On 43 design families written by production coding agents, \sysname{} always picks the same variant as exhaustive synthesis and cuts total evaluation work by 44.9\%. Limitations include evaluation on Design Compiler, the NanGate 45\,nm library, and area targets. Future work will cover other synthesis tools, timing and power objectives, and place-and-route.

\clearpage
\IEEEtriggeratref{23}
\bibliographystyle{IEEEtran}
\bibliography{references}

\end{document}